\documentclass[twocolumn,letterpaper,amsmath,amssymb]{revtex4}
\usepackage{graphicx}
\usepackage{dcolumn}
\usepackage{bm}
\begin{document}

\title{From force distribution to average coordination number \\
  in frictional granular matter}

\author{Ping Wang, Chaoming Song, Christopher Briscoe, Kun Wang,
  Hern\'an A. Makse}

\affiliation {Levich Institute and Physics Department, City College of
  New York, New York, NY 10031, US} \date{\today }

\begin{abstract} {\bf We study the joint probability distribution of
    normal and tangential frictional forces in jammed granular media,
    $P_{\mu}(f_t, f_n)$, for various friction coefficient $\mu$,
    especially when $\mu = \infty$. A universal scaling law is found
    to collapse the data for $\mu=0$ to $\infty$ demonstrating a link
    between force distribution $P_{\mu}(f_t, f_n)$ and average
    coordination number, $z^{\mu}_c$. The results determine $z_c^\mu$
    for a finite friction coefficient, extending the constraints
    counting argument of isostatic granular packing to finite
    frictional packings.}
\end{abstract}

\maketitle

Granular matter undergos the jamming transition evolving
into an amorphous state with a non-zero yield
stress as the density increases to a
point where all particles are in contact \cite{liu}. It has
been shown experimentally and numerically that forces are
inhomogeneously distributed within a jammed granular system, and
further appear to decay exponentially or stretch exponentially for
large values of the force \cite{Radjai,Mueth,Brujic,Silbert}. To date,
there are various theoretical attempts to describe the force
distribution predicting different behavior. For instance, lattice
models and like Boltzmann-equation approaches
\cite{Coppersmith} predict an exponential decay.
Attempts to fit experimental data within the energy ensemble
\cite{Nagel} predict stretched exponential behavior. But the
predictions are difficult to justify, since for granular matter energy
is neither well defined nor conserved due to frictional forces. An
alternative approach is to use the so-called force canonical ensemble
with a Boltzmann distribution where the boundary stress, not energy, is
the conserved quantity \cite{Edwards,Henkes,Rafi}.
It is of interest to reduce the above defined force ensemble to obtain
a single force distribution, but methods to accomplish this remain in
their infancy mainly due to the lack of knowledge on the density of
states \cite{Henkes}. A crude approximation would ignore correlations
between forces and the contact network as well as the density of states
and would predict an exponential decay for the force distribution
\cite{Edwards,Henkes}.

Besides the force distribution and the density of states, an additional
quantity of interest in this study is the average coordination number,
$z^{\mu}_c$, of a system at the jamming transition with interparticle
friction coefficient $\mu$. Despite the importance of $z^{\mu}_c$ for
determining the packing stability, there is only one theoretical
framework to characterize $z^{\mu}_c$ related to the counting argument
of the isostatic conjecture \cite{Alexander}. At the isostatic limit,
the configuration of contact forces has a unique solution if the
contact network is given, since the number of independent forces is
identical to the number of balance equations. Previous works
\cite{Moukarzel,Silbert,Zhang,Kertesz,Shundyak,cph,Snoeijer}
have shown that packings at the jamming transition point are isostatic
\cite{Moukarzel} only for two extreme cases, $\mu = 0$ and
$\mu=\infty$, with average coordination number $z^{0}_c = 2d$ and
$z^{\infty}_c = d+1$ respectively, where $d$ is the dimension. Recent
studies \cite{Kertesz} confirm that the indeterminacy of the force
ensemble \cite{Snoeijer} reaches minimum at $\mu = 0$ and $\infty$.

Lacking more definite theoretical approaches to understand the force
distribution, the density of states and $z^{\mu}_c$ for a general
$\mu$, we perform a numerical study of the joint force
distribution in frictional granular matter, $P_{\mu}(f_t, f_n)$. Here,
the forces at the contacts are normalized by the average forces, in
the tangential direction $f_t=\frac{F_t}{\langle F_t\rangle}$ and in
the normal direction $f_n=\frac{F_n}{\langle F_n\rangle}$. We show
that the key distribution is that of infinite $\mu$, interpreted in
terms of the density of states and exponential statistics, providing
guidance to theoretical attempts under the statistical framework. We
show a universal form of the force ratio distribution $P_{\mu}(u)$,
where $u$ is the ratio of normal and tangential force, $u =
\frac{F_t}{F_n}$ valid for all $\mu$, and a scaling law is found to
collapse all the $P_{\mu}(u)$ determining $z^{\mu}_c$ for
packings. By using $P_{\mu}(u)$ we introduce a way to calculate the
average coordination number for various $\mu$ based on the Maxwell
construction of constraint arguments. Thus, we extend the isostatic
condition from the limits of $\mu=0$ and $\mu=\infty$ to finite $\mu$,
providing the scaling of $z^{\mu}_c$, an unsolved nonlinear
problem. Our results provide a connection between two important
quantities to describe jammed matter: from force distribution to
coordination number.

The packings we studied are composed of 10,000 equal size spheres
interacting with Hertz forces along the contact direction, $F_n$, and
Mindlin forces in the tangential direction, $F_t$, plus the Coulomb
condition, $F_t \leq \mu F_n$ \cite{Zhang}. We first generate a gas
state without friction at an initial volume fraction $\phi_i$, then
the packing is prepared with friction through a slow compression and
relaxation process to achieve equilibrium at a given volume fraction
and coordination number as close as possible to the limiting density
of the jamming transition. A detailed description of the simulation is
given in \cite{cph}.

We start by constructing an empirical formula of $P_{\infty}(f_t,
f_n)$ based on two numerical results:

\begin{figure}
\centering \resizebox{8.5cm}{!}{\includegraphics{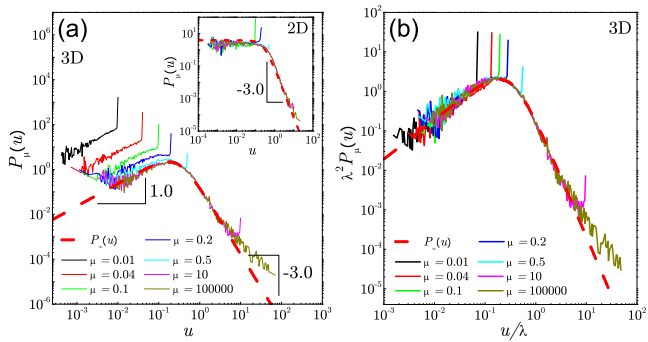}}
\caption{(a) and the inset are log-log plots of the PDF of u
  respectively in 3D and 2D for various $\mu$;
  (b) Log-log plot of the collapsed $P_{\mu}(u)$ for various
  $\mu$ in 3D. The red dash-lines both
  in (a) and (b) are plots of $P_{\infty}(u) = \frac{\kappa(d-1)}{(1+\kappa^2u^2)^{3/2}}(\frac{\kappa u}{\sqrt{1+\kappa^2u^2}})^{d-2}$, where we use $\kappa=3.80$
  and $3.43$ respectively for 2D and 3D from a
  direct measurement of the simulation; }\label{P_u}
\end{figure}

(i) We find that the ratio force distribution
\cite{Radjai,Silbert,Zhang,Shundyak},

\begin{equation}
P_{\mu}(u) = \kappa\int_{0}^{\infty}f_nP_{\mu}(\kappa uf_n, f_n)df_n,
\label{Eq_P_u}
\end{equation}
at infinite friction is characterized by two power-laws with exponents
equal to 0 and -3 in 2D, and 1 and -3 in 3D respectively at
$u\rightarrow 0$ and $u\rightarrow\infty$, where $\kappa =
\frac{\langle F_n\rangle}{\langle F_t\rangle}$, is an anisotropy
parameter. Figure \ref{P_u}a plots $P_{\mu}(u)$ for various values of
$\mu$, showing that all $P_{\mu}(u)$ displays similar behavior having
two power-law slopes except for a sharp peak at $u=\mu$, due to
sliding contacts reaching the Coulomb threshold. A correct form of
force distribution should predict this power-law behavior.

Notice that previous 2D simulations \cite{Radjai,Shundyak}
have reported a plateau of $P_{\mu}(u)$ in the region of $[0, \mu]$,
corresponding to the first power-law of $P_{\mu}(u)$ with the exponent
equal to 0, shown in the inset of Fig. \ref{P_u}a. The second
power-laws only appears for very large values of $\mu$ and has not been
reported by previous studies. We only show $P_{\mu}(u)$ with $\mu >
0.1$ for 2D in the inset of Fig. \ref{P_u}a due to the difficulty of
preparing disordered 2D monodisperse packing at small values of $\mu$.

\begin{figure}
\centering \resizebox{8cm}{!}{\includegraphics{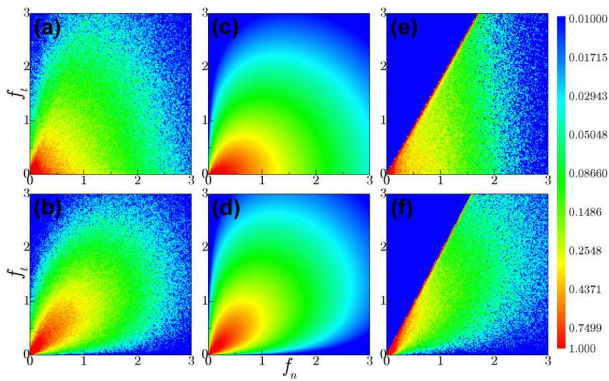}}
\caption{(a), (b) Contour plotting of $P_{\infty}(f_t, f_n)$ from
simulation results in 2D and 3D respectively; (c), (d) Contour
plotting of the empirical formula Eq. (\ref{Eq_P_f_theta}) with
$a=0.8$ in 2D and 3D respectively; (e), (f) Contour plotting of
$P_{0.3}(f_t, f_n)$ from simulation result in 2D and 3D respectively
with $\mu = 0.3$. In (a), (b), (e) and (f), we superpose the data
from $20$ individual configurations, each of them contains 10,000
grains.} \label{Contour}
\end{figure}

(ii) We find that the contour plot of $P_{\infty}(f_t, f_n)$ follows
the geometric behavior shown in Fig. \ref{Contour}a and
\ref{Contour}b, especially in 3D case where $P_{\infty}(f_t,f_n)$ is
symmetric in the space of $(f_t,f_n)$. We will show later on that this
symmetric behavior only occurs at large enough forces in 3D. A correct
form of the force distribution should predict this behavior.



By fitting our numerical data, we find an empirical form of
$P_{\infty}(f_t,f_n)$ for infinite friction, consistent with (i) and
(ii). We describe it by defining new variables $f = \sqrt{f_t^2 +
  f_n^2}$ and $\theta = \arctan(\frac{f_t}{f_n})$,

\begin{equation}
P_{\infty}(f,\theta)= a \,\, g(\theta)\,\,e^{-\sqrt{a} f},
\label{Eq_P_f_theta}
\end{equation}
where $a$ is a constant, which could be regarded as the inverse
of the angoricity \cite{Edwards,Henkes,Rafi}. By fitting this
distribution we find
$$g(\theta)=(d-1)(\sin\theta)^{d-2}\cos\theta,$$ which can be regarded
as the density of states approximately for the force ensemble at
$\mu=\infty$. Equation (\ref{Eq_P_f_theta}), plotted in
Fig. \ref{Contour}c and \ref{Contour}d, shows similar pattern to the
simulation results of Fig. \ref{Contour}a and \ref{Contour}b. We
further study the contour plot of $P_{\mu}(f_t, f_n)$ at $\mu=0.3$
shown in Fig. \ref{Contour}e and \ref{Contour}f. $P_{0.3}(f_t, f_n)$
displays the same pattern inside the Coulomb cone as when
$\mu=\infty$. We therefore suggest that the study of force
distribution for frictional packing should focus on packings with
$\mu=\infty$. The density of state $g(\theta)$ describes the probability of the contact
forces for a single contact to have an angle $\theta$ [we note that
there is no obvious geometric meaning for $\theta$, which is not the
angle between the normal and the net contact force:
$\theta=\arctan(\frac{f_t}{f_n})=\arctan(\kappa\frac{F_t}{F_t})\neq
\arctan(\frac{F_t}{F_t})$] and indicates that normal and tangential
forces are correlated to each other even when there is no Coulomb constraint.

We define $P_<(\theta)$ as the cumulative probability distribution of
$\theta$ indicating the probability of the contact forces for a single
contact to have an angle less than $\theta$, such that
$g(\theta)=\frac{dP_<(\theta)}{d\theta}$. We find:
\begin{equation}
  P_<(\theta)=(\sin\theta)^{d-1}.
\end{equation}
This simple form $P_<$ could lead to a theoretical approach to the
force distribution since it provides the density of states within the statistical mechanics framework
\cite{Henkes,Rafi}.

\begin{figure}
\centering \resizebox{8cm}{!}{\includegraphics{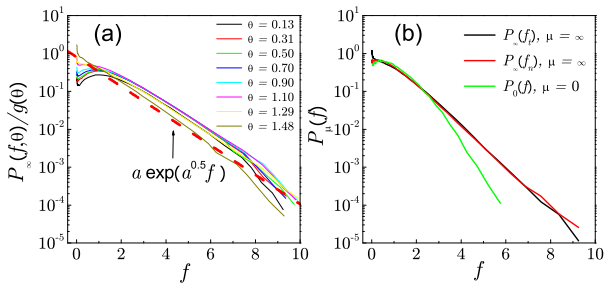}}
\caption{
(a) Log-linear plot of $P_{\infty}(f,\theta)/g(\theta)$ for various
value of $\theta$ in 3D. All the curves well collapse with a pure
exponential tail in the region of $f>1$. The red dashed-line is
function of $ a e^{-\sqrt{a} f}$ with $a = 0.8$. (b)
Log-linear plot of $P_{\infty}(f_t)$, $P_{\infty}(f_n)$ and
$P_{0}(f_n)$.} \label{P_f_theta}
\end{figure}

Equation (\ref{Eq_P_f_theta}) implies that $P_\infty(f,
\theta)/g(\theta) = a e^{-\sqrt{a} f}$ is independent of
$\theta$. We plot $P_\infty(f, \theta)/g(\theta)$ for various values
of $\theta$ in Fig. \ref{P_f_theta}a to further
compare with the simulation results. We find that all the curves
collapse with exponential tails in the region of $f>1$, indicating
that the empirical form of Eq. (\ref{Eq_P_f_theta}) captures the main
features of the force distribution for large forces. $P_\infty(f,
\theta)/g(\theta)$ has a peak at $f\simeq1$ when $\theta$ is small,
and exhibits a monotonic exponential decrease when $\theta$ is close
to $\frac{\pi}{2}$. This implies that the probabilities of single
forces, $P_{\infty}(f_n)$ and $P_{\infty}(f_t)$, have different
behavior as shown in Fig. \ref{P_f_theta}b:
$P_{\infty}(f_n)$ displays a peak at $f\simeq1$ while
$P_{\infty}(f_t)$ does not. This result is consistent with previous
experimental studies of frictional packings \cite{Behringer}.
In Fig. \ref{P_f_theta}b we plot the force
distribution at $\mu=0$ and compare with $\mu=\infty$.  We conclude
that $P_{0}(f_n)$ has a stretched exponential tail close to Gaussian with a
exponent $\beta=1.65$ due to local entropy maximization \cite{Tighe}.

By using Eq. (\ref{Eq_P_u}) and Eq. (\ref{Eq_P_f_theta}) we obtain a
ratio force distribution $P_{\infty}(u)$, shown
in Fig. \ref{P_u}, as a red dashed-line
in good agreement with numerical results. This result further confirms
that our empirical formula Eq. (\ref{Eq_P_f_theta}) is reasonable.

Further, we show that the ratio distribution is the link
between the ensemble of forces and the average coordination number. We
find that $P_\mu(u)$ can be rescaled to a single curve (except for the
peak at $\mu$), with scaling factors equal to $\lambda$ and
$\lambda^2$, for the $y$ and $x$ axes, respectively. We find
$\lambda=1$ in 2D and $\lambda =
(z^0_c-z^{\infty}_c)/(z^0_c-z^{\mu}_c)$ in 3D, as plotted in
Fig. \ref{P_u}b. In the 2D case, $P_{\mu}(u)$ collapses without
scaling, so $\lambda=1$. The 3D case is different, where we find that
$\lambda\to 1$ when $\mu\to \infty$, so $P_{\infty}(u)$ does not
change after multiplying the scaling factors. The factor $\lambda$
diverges at $\mu=0$, implying that $P_{\mu}(u)$ reduces to a delta
function at $\mu=0$ due to the fact that all contact forces reach the
Coulomb threshold in a pure frictionless packing.

Next, we show that the universal form of $P_\infty(u)$ determines
$z^{\mu}_c$ for any $\mu$, hereby extending the isostatic counting
argument from $\mu = 0$ and $\mu = \infty$ to finite values of
$\mu$. From linear counting arguments we know that $z^{0}_c = 2d$ and
$z^{\infty}_c = d+1$, and we want to interpolate to finite $\mu$ and
obtain $z^{\mu}_c$.
Below, we show that the Maxwell constraint arguments based on
the number of redundant constraints provides the framework to derive
$z^{\mu}_c$. Analysis of the coordination number of granular packings
can be related to the Maxwell constraints counting in the rigidity
percolation theory \cite{Maxwell}:
\begin{equation}
F = \frac{zd}{2}N - N_c + N_r,
\label{maxwell}
\end{equation}
where $F$ is the number of degrees of freedom (or floppy modes)
satisfying $F\geq 0$, $N$ is the number of grains, $N_c$ is the number
of constraints, $N_r$ is the number of redundant constraints, and $z$
is the coordination number. At the jamming transition, $F=0$, resulting
in a minimum value of $z$, i.e., $z^{\mu}_c$. Here $zdN/2$ is equal to
the total number of unknown force variables for a fixed force network.


We consider a static packing with both force and torque balances, but
without any typical constraints of translation and rotation. For
packings with $\mu = \infty$, the number of constraint $N_c$ will be
equal to the number of force balance equation, $dN$, plus the number
of torque balance equation, $d(d-1)N/2$, i.e., $N_c(\infty) =
d(d+1)N/2$. There exists reasonable evidence
\cite{liu,Silbert,Moukarzel,Kertesz,Snoeijer,Alexander,Zhang,Shundyak,cph}
to believe that at the jamming transition, $N_r(\infty) = 0$, implying
a conjecture that the Maxwell counting approximation is
exact. Therefore, $z = z^{\infty}_c = d+1$. Another important case is
at $\mu = 0$. Here the redundant constraints, $N_r(0) = d(d-1)N/2$, is
equal to the number of torque balance equation due to the absence of
tangential force. Further, we must add $z(d-1)N/2$ extra constraints
to $N_c(0)$, corresponding to equations of tangential force equal to
zero, ${F_t}^i = 0$. Therefore, $N_c(0) = N_c(\infty) + z(d-1)N/2$ and
we obtain $z = z^0_c = 2d$.

Analyzing intermediate values of $\mu$, is complicated since many
inequality constrains are created as $\mu {F_n}^i - {F_t}^i \geq
0$. Calculating $z^{\mu}_c$ becomes a nonlinear problem and can be
understood as an optimization of an outcome based on some set of
constraints, i.e., minimizing a Hamiltonian of the system,
$\mathbb{H}(\mathbf{F_n}, \mathbf{F_t})$, over a convex polyhedron
specified by linear and non-negativity constraints. An interesting
feature found in previous studies is that $z^{\mu}_c$ monotonically
decreases from $2d$ to $d+1$ with increasing $\mu$
\cite{Silbert,Kertesz,Shundyak,cph}, implying that we can map this
non-linear problem to a linear one by considering a monotonic change
in the number of constraints in Eq. (\ref{maxwell}) with increasing
$\mu$.

The above analysis suggests to extend the Maxwell counting argument
Eq. (\ref{maxwell}) to a system with finite $\mu$ as:

\begin{equation}
F = \frac{z^{\mu}_cd}{2}N - N_c(\infty) + \left[N_r(0)- z^{\mu}_c(d-1)N/2\right]\eta(\mu) = 0,
\label{maxwell2}
\end{equation}
where $\eta(\mu)$ is an undetermined monotonic function ranging from
$1$ to $0$ as $\mu$ ranges from $0$ to $\infty$. The problem is
reduced to choosing a functional form for $\eta(\mu)$.


To determine $\eta(\mu)$, we notice that it should be related to the
sliding rate of packings, i.e., the ratio of the number of the sliding
contacts to the number of total contacts in a packing, denoted
$S(\mu)$.  By definition, $S(\mu)$ is determined by $P_\mu(u)$,
providing a link between coordination number an force distribution:

\begin{equation}
S(\mu)=1-{\int_{0}}^{\mu}P_{\mu}(u)du=1-{\int_{0}}^{\mu}\lambda^2P_{\infty}(\lambda u)du
\label{Eq_S_mu}
\end{equation}
The limiting cases are $S(0)=1$ and $S(\infty)=0$, and $S(\mu)$ has
the same monotonic behaviour as $\eta(\mu)$.

While $\eta(\mu)$ must be a function of $S(\mu)$, there are many
choices for the functional relation between both quantities.
We determine this functional form by fitting the simulations. Setting $\eta(\mu)=1-(1-S(\mu))/\lambda$ provides very good fitting of
$z^{\mu}_c$ with simulations both in 2D and 3D. Substituting $\eta(\mu)$ into
Eq. (\ref{maxwell2}), we arrive at a cubic equation for $z^{\mu}_c$ in 3D:

\begin{equation}
  \frac{1}{\kappa^2\mu^2}\left(\frac{6-z^{\mu}_c}{2}\right)^3+3\left(\frac{6-z^{\mu}_c}{2}\right)-3=0.
\label{cubic}
\end{equation}

\begin{figure}
\centering \resizebox{8cm}{!}{\includegraphics{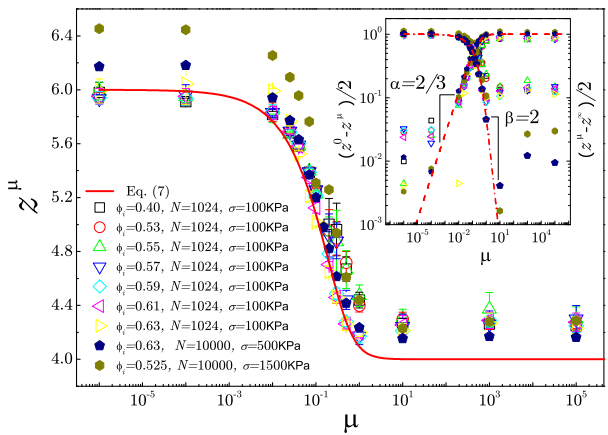}}
\caption{$z^\mu$ versus $\mu$ for various initial volume fraction
$\phi_i$ in 3D. The red solid line is the theory result predicted
by Eq. (\ref{cubic}). In the inset, the red-dash and dash-dot lines
are the prediction of $(z^{0}-z^{\mu})/2$ and $(z^{\mu}-z^{\infty})/2$,
respectively, in comparison with simulations.} \label{zc_mu}
\end{figure}

It can be shown that Eq. (\ref{cubic}) predicts two power-law
relations, $z^{0}_c-z^{\mu}_c\sim\mu^\alpha$, and
$z^{\mu}_c-z^{\infty}_c\sim\mu^{-\beta}$,
respectively for $\mu\rightarrow0$ and $\mu\rightarrow\infty$, where
$\alpha=2/3$ and $\beta=2$.
In Fig. \ref{zc_mu}
we plot $z^{\mu}_c$ obtained from the cubic Eq. (\ref{cubic}) and
compare with simulation data in 3D. The asymptotic predictions of
$\alpha=2/3$ and $\beta=2$ are in good agreement with simulation
results shown in the inset of Fig. \ref{zc_mu}. It is difficult to
check the value of $\beta$ due to the difficulty of preparing a 3D
packing as close as possible to $z^{\infty}_c=4$. To solve this
problem, we prepare larger packings slightly above the critical point
with a small constant pressure, and $z^{\mu}_c$ is replaced by $z^{\mu}$
without suffix. This result is shown in Fig. \ref{zc_mu} with two sets
of data for pressure $\sigma=500\mathrm{Kpa}$ and $\sigma=1500\mathrm{KPa}$.
We can see that the power law of coordination number is independent of
pressure even when $z^{\mu}$ is far from the isostatic value.

When we combine power-law finding of $z^{\mu}_c$ with our
theoretical work of \cite{cph} in 3D, where $z^{\mu}_c$ is linked to
the volume fraction $\phi^{\mu}_c$ with a simple formula,
$\phi^{\mu}_c=z^{\mu}_c/(z^{\mu}_c+2\sqrt{3})$,
then we solve the relation between $\phi^{\mu}_c$, $z^{\mu}_c$ and
$\mu$; $\phi^{\mu}_c$ follows the same scaling behavior with $\mu$,
$\phi^{0}_c-\phi^{\mu}_c\sim\mu^{\alpha}$, and $\phi^{\mu}_c-\phi^{\infty}_c\sim\mu^{-\beta}$. Recent experiments
\cite{Jerkins} in 3D investigate the preparation of packings close to
the random loose packing limit. They find $\alpha = 0.51\pm0.25$ and
$\beta = 0.89\pm0.16$. Their measurement of $\beta$ is far away from
our prediction which could be due to the same reason as us, i.e., the
difficulty of preparing packing as close as possible to $z^{\infty}_c=4$.

In the 2D case, $\eta(\mu)=S(\mu)$ since $\lambda=1$, and we have

\begin{equation}
  z^{\mu}_c=\left[4+\frac{2\kappa\mu}{(1+\kappa^2\mu^2)^{1/2}}\right]/\left[1+\frac{\kappa\mu}{(1+\kappa^2\mu^2)^{1/2}}\right].
\end{equation}
This equation predicts $\alpha=1$ and $\beta=2$, close to our
simulation result of $\beta=1.86$. We can not determine the value of
$\alpha$ from simulation due to the difficulty of preparing disordered
2D monodisperse packings for small values of $\mu$, and the
polydispersity of packings may slightly affect these two
indices. Previous simulations \cite{Shundyak} of polydisperse 2D
packings have $\alpha=0.7$, still close to our predictions.

In summary, we develop a framework
to study the connection between the force
distribution and the coordination number.
Some
aspects of this connection remain empirical, including the density of
states, $g(\theta)$, and the scaling factor $\lambda$, allowing for
the collapse of $P_{\mu}(u)$ into a single curve.
Overall, the obtained mathematical forms of the density of states, the
different force distributions, the coordination number and volume
fraction, may allow for their incorporation into a statistical force
ensemble of jammed matter \cite{Henkes,Rafi}. This may facilitate the
solution of outstanding open problems such as the prediction of the
power law scaling of the pressure, the coordination number and elastic
moduli with the volume fraction near the jamming transition
\cite{liu}.

\end{document}